\documentclass[10pt,letter]{article}                                                                   
\usepackage{graphicx}
\usepackage{amssymb}
\newcommand{\ud}{\mathrm{d}}

\begin{document}                                                                                    
\thispagestyle{empty}                                                              
                                                                                   
\begin{center}                                                                     
\begin{tabular}{p{130mm}}                                                          
                                                                                   
\begin{center}                                                                     
{\bf\Large                                                                         
LOCALIZATION AND FUSION MODELING }\\
\vspace{5mm}  

{\bf\Large IN PLASMA PHYSICS.} \\                                                  
\vspace{5mm}                                                                       
                                                                                   
{\bf\Large  PART I: MATH FRAMEWORK FOR }\\                                         
\vspace{5mm}                                                                       
                                                                                   
{\bf\Large NON-EQUILIBRIUM HIERARCHIES}\\                                                      
                                                                                   
\vspace{1cm}                                                                       
                                                                                   
{\bf\Large Antonina N. Fedorova, Michael G. Zeitlin}                             
                                                                                   
\vspace{1cm}

{\bf\large\it                                                                      
IPME RAS, St.~Petersburg,                                                          
V.O. Bolshoj pr., 61, 199178, Russia}\\                                            
{\bf\large\it e-mail: zeitlin@math.ipme.ru}\\                                      
{\bf\large\it e-mail: anton@math.ipme.ru}\\                                        
{\bf\large\it http://www.ipme.ru/zeitlin.html}\\                                   
{\bf\large\it http://www.ipme.nw.ru/zeitlin.html}                                
                                                                                   
\end{center}                                                                       
                                                                                   
\vspace{2cm}                                                                       
                                                                                   
\begin{abstract}                                                                   
A fast and efficient numerical-analytical approach is proposed for                                
description of complex behaviour in non-equilibrium ensembles in the                           
BBGKY framework. We construct the multiscale representation for                                   
hierarchy of partition functions by means of the variational approach                             
and multiresolution decomposition. Numerical modeling shows the creation                          
of various internal structures from fundamental localized (eigen)modes.                           
These patterns determine the behaviour of plasma. The localized pattern                           
(waveleton) is a model for energy confinement state (fusion) in plasma.
\end{abstract}

\vspace{10mm}                                                                      
                                                                                   
\begin{center}                                                                     
{\large 
Two lectures presented at the Sixth Symposium on
Current Trends in International Fusion Research, Washington D.C., March, 2005,}\\      
        
{\large edited by Emilio Panarella, NRC Reasearch Press, National Reasearch Council of Canada,           
Ottawa, Canada, 2006.}                                                     
\end{center}
                                                                       
\end{tabular}                                                                      
\end{center}


\newpage

\title{
LOCALIZATION AND FUSION MODELING IN PLASMA PHYSICS. PART I: MATH
FRAMEWORK FOR NON-EQUILIBRIUM HIERARCHIES\footnote{
Current Trends in International Fusion Research - Proceedings of the Sixth Symposium
Edited by Emilio Panarella. NRC Reasearch Press, National Reasearch Council of Canada, 
Ottawa, ON K1A 0R6 Canada, 2006.}}

\author{Antonina~ N. Fedorova and Michael~ G. Zeitlin \\
IPME RAS, Russian Academy of Sciences, \\
V.O. Bolshoj pr., 61, \\
199178, St. Petersburg, Russia \\
http://www.ipme.ru/zeitlin.html;\\
http://www.ipme.nw.ru/zeitlin.html 
}
\date{}
\maketitle
\thispagestyle{empty}

\begin{abstract}
A fast and efficient numerical-analytical approach is proposed for
description of complex behaviour in non-equilibrium ensembles in the\\
BBGKY framework. We construct the multiscale representation for
hierarchy of partition functions by means of the variational approach
and multiresolution decomposition. Numerical modeling shows the creation
of various internal structures from fundamental localized (eigen)modes.
These patterns determine the behaviour of plasma. The localized pattern
(waveleton) is a model for energy confinement state (fusion) in plasma.
\end{abstract}

\begin{flushright}
{\bf ``A magnetically confined plasma cannot \\
be in thermodinamical equilibrium'' }\\
Unknown author ... Folklore \\
\end{flushright}

\section{GENERAL INTRODUCTION}
It is well known that fusion problem in plasma physics could be solved neither 
experimentally nor theoretically during last fifty years. At the same time, during this long 
period other areas of physics and engineering demonstrated vast growth, on the level of both 
theoretical understanding and practical smart realizability. We can only mention an 
unprecedent level of theoretical understanding in Quantum Field Theory and String Physics, 
as the top of this mountain, and solid state electronics penetrating in real life together with 
personal computers and a lot of related things. It should be mentioned that the former thing 
demanded and created a fantastic level of theoretical models and new beatiful mathematics 
while the latter depended only on the state of the art of engineers from Intel and 
other high-tech firms who created, e.g., the processor Pentium 
by means of the multiplication table 
(almost) only. Unfortunately, although plasma physics, as a whole, was a key source of the so
 called Soliton Theory (begining with numerical modeling by M. Kruskal
and N. Zabusky), which was very important part of Mathematical Physics
during twenty-five years, it itself used no advantage of these methods
and approaches and created nothing comparable for a long time.

Because financing contributions in this area definitely exceeds that of
almost all other areas of Physics, it seems that there are the serious
obstacles which prevent real progress in the problem of real fusion as
the main subject in the area. Of course, it may be a result of some
unknown no-go theorem but it seems that the current theoretical level
definitely demonstrates that not all possibilities, at least on the
level of theoretical and matematical modeling, are exhausted. Surely, it
is more than clear that perturbations, linearization, PIC or MC do not
exhaust all instruments which we have at our hands on the route to
theoretical understanding and predictions. Definitely, we need much more
to have influence on almost free-of-theoretical-background work of
experimentalists and engineers who contribute to ITER and other related
top level projects. 

So, this paper and related ones can be considered as a small
contribution to an attempt to avoid the existing obstacles appeared on
the main road of current plasma physics. 

Definitely, the first thing which we need to change is a framework of
generic mathematical methods which can only improve the current state of
the theory. 

Although, as in case of Pentium processors, there is a chance that
engineers can solve all problems in a smart way without any theoretical
contributions, but the current status of out-of-Science problems is such
one that, definetely, we have no additional fifty years to wait when the
appearance of fusion machines leads to total decreasing both the level
of using of oil and the level of prices adequate for civilized
countries. 

So, after pointing out our vision of well known situation in the area,
we will present some details of our attempts to change the standard
approach [1], [2] to more proper one from our point of view [3]-[8]. 

Our postulates (conjectures) are as follows: 

{\bf A)} The fusion problem (at least at the first step) must be considered as
a problem inside the (non) equilibrium ensemble in the full phase space.
It means, at least, that: 

{\bf A1)} our dynamical variables are partitions (partition functions,
hierarchy of N-point partition functions), 

{\bf A2)} it is impossible to fix a priori the concrete distribution function
and postulate it (e.g. Maxwell-like or other concrete distributions)
but, on the contrary, the proper distrubution(s) must be the solutions
of proper (stochastic) dynamical problem(s), e.g., it may be the
well-known framework of BBGKY hierarchy of kinetic equations or
something similar. So, the full set of dynamical variables must include
partitions also. 

{\bf B)} Fusion state = (meta) stable state (in the space of partitions on the
whole phase space) in which most of energy of the system is concentrated
in the relatively small area (preferable with measure zero) of the whole
domain of definition in the phase space during the time period which is
enough to take reasonable part of it outside for possible usage. 

From the formal/mathematical point of view it means that: 

{\bf B1)} fusion state must be localized (first of all, in the phase space), 

{\bf B2)} we need a set of building blocks, localized basic states or
eigenmodes which can provide 

{\bf B3)} the creation of localized pattern which can be considered as a
possible model for plasma in a fusion state. Such pattern must be: 

{\bf B4)} (meta) stable and controllable because of obvious reasons. 

So, in this paper our main courses are: 

{\bf C1)} to present smart localized building blocks which may be not only
useful from point of view of analytical statements, such as the best
possible localization, fast convergence, sparse operators representation,
etc [9], but also exist as real physical fundamental modes, 

{\bf C2)} to construct various possible patterns with special attention to
localized pattern which can be considered as a needful thing in analysis
of fusion; 

{\bf C3)} after points C1 and C2 in ensemble (BBGKY) framework to consider
some standard reductions to Vlasov-like and RMS-like equations
(following the set-up from well-known results [2]) which may be useful
also. These particular cases may be important as from physical point of
view as some illustration of general consideration [10]. 

Sections 5 (Part I), and 1, 2 (Part II [10]) are more or less
self-consistent, so potential readers may start from any of these
Sections.

The lines above are motivated by our attempts to analyze the hidden
internal contents of the phrase mentioned in the epigraph of this paper:
``A magnetically confined plasma cannot be in thermodinamical
equilibrium.'' We guess that it is a well-known and well distributed
fact. We found it, in particular, in the recent
review paper [1]. Of course, the following is only a first order
iteration on the long way to the main goal but we think that one needs
to start from the right place to have some chance to reach the final
point.

\section{MOTIVATIONS: INTRODUCTION INTO\\
 MATHEMATICAL FRAMEWORK.\\
 KINDERGARTEN VERSION}

It is obvious that any reasonable set-up for analysis of fusion leads to
very complex system and one hardly believes that such system can be
analyzed by means of almost exhausted methods like perturbations,
linearization etc. At the same time, because such stochastic dynamics is
overcompleted by short- and long-living fluctuations, instabilities, etc
one needs to find something more proper than usual plane waves or
gaussians for modeling a complicated complex behaviour. For
reminiscences we may consider simple standard soliton equations like
KdV, KP or sine-Gordon ones. It is well-known that neither
linearization, nor perturbations, nor plane-wave-like approximations are
proper for the reasonable analysis of such equations in contrary to wave
or other simple linear equations: it is impossible to approximate the
spectrum of such models (solitons, breathers or finite-gap solutions) by
means of linear Fourier harmonics. They are not physical modes in
complex situation. Moreover, such linear methods are not adequate in
more complicated situation. 

Below, we will give kindergarten-like introduction to our analytical
framework. It would appear that as a first step in this direction is to
find a reasonable extension of understanding of the non-equilibrium
dynamics as a whole. One needs to sketch up the underlying ingredients
of the theory (spaces of states, observables, measures, classes of
smoothness, dynamical set-up etc) in an attempt to provide the maximally
extendable but at the same time really calculable and realizable
description of the complex dynamics inside hierarchies like BBKGY and
their reductions. The general idea is rather simple: it is well known
that the idea of ``symmetry'' is the key ingredient of any reasonable
physical theory from classical (in)finite dimensional (integrable)
Hamiltonian dynamics to different sub-planckian models based on strings
(branes, orbifolds, etc). During the last century kinematical, dynamical
and hidden symmetries played the key role in our understanding of
physical process. Roughly speaking, the representation theory of
underlying symmetry (classical or quantum, groups or (bi)algebras,
finite or infinite dimensional, continuous or discrete) is a proper
instrument for the description of proper (orbital) dynamics. A starting
point for us is a possible model for fundamental localized mode with the
subsequent description of the whole zoo of possible realizable
(controllable) states/patterns which may be useful from the point of
view of experimentalists and engineers. The proper representation theory
is well known as ``local nonlinear harmonic analysis'', in particular
case of simple underlying symmetry-affine group-aka wavelet analysis.
From our point of view the advantages of such approach are as follows: 

\noindent{\bf i)} natural realization of localized states in any proper functional
realization of (functional) space of states, 

\noindent{\bf ii)} hidden symmetry of chosen realization of proper functional model
provides the (whole) spectrum of possible states via the so-called
multiresolution decomposition. 

So, indeed, the hidden symmetry (non-abelian affine group in the
simplest case) of the space of states via proper representation theory
generates the physical spectrum and this procedure depends on the choice
of the functional realization of the space of states. It explicitly
demonstrates that the structure and properties of the functional
realization of the space of states are the natural properties of
physical world at the same level of importance as a particular choice of
Hamiltonian, or the equation of motion, or the action principle
(variational method). 

At the next step we need to consider the consequences of our choice i),
ii) for the algebra of operators. In this direction one needs to mention
the class of operators we are interested in to present the proper
description for a maximally generalized but reasonable class of
problems. It seems that these must be pseudodifferential operators,
especially if we underline that in the spirit of points i), ii) above we
need to take Weyl framework for analyzing basic equations of motion. It
is obvious, that the consideration of symbols of operators instead of
operators themselves is the starting point as for the correct
mathematical theory of pseudodifferential operators as for analysis of
dynamics formulated in the language of symbols (Wigner-Weyl transform).
In addition, it provides us by unified framework for analysis as
classical problems as quantum ones.

It should be noted that in such picture 
we can naturally include 
the effects of self-interaction on the way 
of construction and subsequent analysis of nonlinear models. So, our 
consideration will be in the framework of (Nonlinear) Pseudodifferential Dynamics  ($\Psi DOD$).
 As a result of i), ii), we will have: 

\noindent{\bf iii)} most sparse, almost diagonal, representation for a wide class of
operators included in the set-up of the whole problems. 

It is possible by using the so-called Fast Wavelet Transform
representation for algebra of observables. 

Then points i)-iii) provide us by 

\noindent{\bf iv)} natural (non-perturbative) multiscale decomposition for all
dynamical quantities such as states, observables, partitions. 

Existence of such internal multiscales with different dynamics at each
scale and transitions, interactions, and intermittency between scales
demonstrates that statistical mechanics in BBGKY form, despite its
linear structure, is really a complicated problem from the mathematical
point of view. It seems, that well-known underlying ``stochastic''
complexity is a result of transition by means of (still rather unclear)
procedure of evolution from complexity related to individual classical
dynamics to the rich pseudodifferential (more exactly, microlocal)
structure on the non-equilibrium ensemble side. 

We divide all possible configurations related to possible solutions of
our kinetic hierarchies into two classes: 

{\bf (a)} standard solutions; {\bf (b)} controllable solutions (solutions with
prescribed qualitative type of behaviour). 

Anyway, the whole zoo of solutions consists of possible patterns,
including very important ones from the point of view of underlying
physics: 

\noindent{\bf v)} localized modes (basis modes, eigenmodes) and chaotic (definitely,
non pro\-per for fusion modeling) or fusion-like localized patterns
constructed from them. 

It should be noted that these bases modes are nonlinear in contrast with usual ones 
because they come from (non) abelian generic group while the usual Fourier (commutative) 
analysis starts from   $U(1)$
 abelian modes (plane waves). They are really ``eigenmodes'' but in
sense of decomposition of representation of the underlying hidden
symmetry group which generates the multiresolution decomposition. The
set of patterns is built from these modes by means of variational
procedures more or less standard in mathematical physics. It allows to
control the convergence from one side but, what is more important, 

\noindent{\bf vi)} to consider the problem of the control of patterns (types of
behaviour) on the level of reduced (variational) algebraical equations. 

We need to mention that it is possible to change the simplest generic
group of hidden internal symmetry from the affine (translations and
dilations) to much more general, but, in any case, this generic symmetry
will produce the proper natural high localized eigenmodes, as well as
the decomposition of the functional realization of space of states into
the proper orbits; and all that allows to compute dynamical consequence
of this procedure, i.e. pattern formation, and, as a result, to classify
the whole spectrum of proper states. 

For practical reasons controllable patterns (with prescribed behaviour)
are the most useful. We mention the so-called waveleton-like pattern
which we regard as the most important one. We use the following allusion
in the space of words: 

\{waveleton\}:=\{soliton\}  $\bigsqcup$   \{wavelet\}

It means: 

\noindent{\bf vii)} waveleton $\approx$ (meta)stable localized (controllable) pattern. 

To summarize, the approach described below allows 

\noindent{\bf viii)} to solve wide classes of general $\Psi DOD$
 problems, including generic for us BBGKY hierarchy and its reductions,
and 

\noindent{\bf ix)} to present the analytical/numerical realization for physically
interesting patterns, like fusion states. 

We would like to emphasize the effectiveness of numerical realization of
this program (minimal complexity of calculations) as additional
advantage. So, items i)-ix) point out all main features of our approach
[3]-[8].

\subsection{Class of Models}

Here we describe a class of problems which can be analyzed by methods
described in the previous part. We start from individual dynamics and
finish by (non)-equilibrium ensembles. Me mention here as classical
models as quantum ones because both of them can be immersed in this
unified framework and it is really motivated by ``quantum-like''
ideology. 

All models belong to the $\Psi DOD$ class and can be described by a finite or infinite 
(named hierarchies in such cases) system of $\Psi DOD$
 equations:

\noindent{\bf a)} Individual classical/quantum mechanics ($cM/qM$): linear/nonlinear; 
$\{cM\}\subset\{qM\}$, $\ast$  - quantized for the class of polynomial Hamiltonians
$
H(p,q,t)=\sum_{i.j}a_{ij}(t)p^iq^j.
$

An important generic example: Orbital motion (in Storage Rings).
The magnetic vector potential of a magnet with $2n$
 poles in Cartesian coordinates is 
\begin{equation}
A=\sum_n K_nf_n(x,y),
\end{equation}
	
where  $f_n$  is a homogeneous function of $x$ and  $y$  of order  $n$. The cases  $n=2$ to $n=5$ 
 correspond to low-order multipoles: quadrupole, sextupole, octupole,
decupole. The corresponding Hamiltonian is: 

\begin{eqnarray}\label{eq:ham}
H(x,p_x,y,p_y,s)&=&\frac{p_x^2+p_y^2}{2}+
\left(\frac{1}{\rho^2(s)}-k_1(s)\right)
\cdot\frac{x^2}{2}+k_1(s)\frac{y^2}{2}\nonumber\\
&-&{\cal R}e\left[\sum_{n\geq 2}
\frac{k_n(s)+ij_n(s)}{(n+1)!}\cdot(x+iy)^{(n+1)}\right].
\end{eqnarray}	

Terms corresponding to kick type contributions of rf-cavity: 

\begin{eqnarray}
A_\tau=-\frac{L}{2\pi k}\cdot V_0\cdot \cos\big(k\frac{2\pi}
 {L}\tau\big)\cdot\delta(s-s_0)
\end{eqnarray}		
or localized cavity   with $V(s)=V_0\cdot \delta_p(s-s_0)$ with $\delta_p(s-s_0)=
\sum^{n=+\infty}_{n=-\infty}\delta(s-(s_0+n\cdot L))$
  at position  $s_0$.
The second example: using Serret-Frenet parametrization, after
truncation of power series expansion of square root we have the
following approximated (up to octupoles) Hamiltonian for orbital motion
in machine coordinates:
\begin{eqnarray}
{\cal H}&=&
   \frac{1}{2}\cdot\frac{[p_x+H\cdot z]^2 + [p_z-H\cdot x]^2}
{[1+f(p_\sigma)]}
+p_\sigma-[1+K_x\cdot x+K_z\cdot z]\cdot f(p_\sigma)\nonumber\\
&+&\frac{1}{2}\cdot[K_x^2+g]\cdot x^2+\frac{1}{2}\cdot[K_z^2-g]\cdot z^2-
  N\cdot xz 
+\frac{\lambda}{6}\cdot(x^3-3xz^2)\\
&+&\frac{\mu}{24}\cdot(z^4-6x^2z^2+x^4)
+\frac{1}{\beta_0^2}\cdot\frac{L}{2\pi\cdot h}\cdot\frac{eV(s)}{E_0}\cdot
\cos\left[h\cdot\frac{2\pi}{L}\cdot\sigma+\varphi\right]\nonumber
\end{eqnarray}
Then we use series expansion of function $f(p_\sigma)$:
$
f(p_\sigma)=f(0)+f^\prime(0)p_\sigma+f^{\prime\prime}(0)
p_\sigma^2/2+\ldots
=p_\sigma-
p_\sigma^2/(2\gamma_0^2)+\ldots$

and the corresponding expansion of RHS of equations. As a result,
important models (2) and (4) belong to a class of polynomial
Hamiltonians, generic for us class of individual dynamics.

\noindent{\bf b)} QFT-like models in framework of the second quantization (dynamics in
Fock spaces). 

\noindent{\bf c)} Classical (non) equilibrium ensembles via BBGKY Hierarchy (with
reductions to different forms of Vlasov-Maxwell/Poisson equations). 

\noindent{\bf d)} Wignerization of a): Wigner-Moyal-Weyl-von Neumann-Lindblad. 

\noindent{\bf e)} Wignerization of c): Quantum (Non) Equilibrium Ensembles.

Important remarks: points a)-e) are considered in  $\Psi DO$ picture of (Non)Linear $\Psi DO$
 Dynamics (surely, all $qM\subset\Psi DOD$);
 dynamical variables/observables are the symbols of operators or
functions; in case of ensembles, the main set of dynamical variables
consists of partitions (n-particle partition functions).

\subsection{Effects we are interested in}
\begin{itemize}
\item[i)] Hierarchy of internal/hidden scales (time, space, phase space). 

\item[ii)] Non-perturbative multiscales: from slow to fast contributions, from
the coarser level of resolution/decomposition to the finer one. 

\item[iii)] Coexistence of hierarchy of multiscale dynamics with transitions
between scales. 

\item[iv)] Realization of the key features of the complex nonequilibrium world
such as the existence of chaotic and/or entangled (complex) states with
possible destruction in some controllable regimes and transition to
localized fusion-like states. 
\end{itemize}

At this level we may interpret the effect of intermode interactions (in
linear but pseudodifferential system!) as a result of simple interscale
interaction or intermittency (with allusion to hydrodynamics), i.e. the
mixing of orbits generated by multiresolution representation of hidden
underlying symmetry. Surely, the concrete realization of such a symmetry
is a natural physical property of the physical model as well as the
space of representation and its proper functional realization. 

One additional important comment: as usual in modern physics, we have
the hierarchy of underlying symmetries; so our internal symmetry of
functional realization of space of states is really not more than
kinematical, because much more rich algebraic structure, related to
operator Cuntz algebra and quantum groups, is hidden inside. The proper
representations can generate much more interesting effects than ones
described above. We will consider it elsewhere but mention here only how
it can be realized by the existing functorial maps between proper
categories: 

\{QMF\} $\longrightarrow$ Loop groups $\longrightarrow$ Cuntz operator algebra 
$\longrightarrow$ Quantum Group 
structure, where \{QMF\} are the so-called quadratic mirror filters generating the realization of 
multiresolution decomposition/representation in any functional space; loop group is well 
known in many areas of physics, e.g. soliton theory, strings etc, roughly speaking, its algebra 
coincides with Virasoro algebra; Cuntz operator algebra is universal $C^*$
 algebra generated by N elements with two relations between them;
Quantum group structure (bialgebra, Hopf algebra, etc) is well known in
many areas because of its universality. It should be noted the
appearance of natural Fock structure inside this functorial sequence
above with the creation operator realized as some generalization of
Cuntz-Toeplitz isometries. Surely, all that can open a new vision of old
problems and bring new possibilities. 

We finish this part by the following qualitative definitions of key
objects (patterns). Their description and understanding in various
physical models is our main goal in this direction. 

\begin{itemize}
\item
By localized states (localized modes) we mean the building blocks for
solutions or generating modes which are localized in maximally small
region of the phase (as in c- as in q-case) space. 
\item
By a chaotic pattern we mean some solution (or asymptotics of solution)
which has random-like distributed energy spectrum in the full domain of
definition. 
\item
By a localized pattern (waveleton) we mean (asymptotically) (meta)
stable solution localized in a relatively small region of the whole
phase space (or a domain of definition). In this case the energy is
distributed during some time (sufficiently large) between only a few
localized modes (from point 1). We believe it to be a good model for
plasma in fusion state (energy confinement).
\end{itemize}

\subsection{Methods}

\begin{itemize}
\item[\bf i)] Representation theory of internal/hidden/underlying symmetry,
Kinematical, Dynamical, Hidden. 

\item[\bf ii)] Arena (space of representation): proper functional realization of
(Hilbert) space of states. 

\item[\bf iii)] Harmonic analysis on (non)abelian group of internal symmetry. Local
Nonlinear (non-abelian) Harmonic Analysis (e.g, wavelet/gabor etc.
analysis) instead of linear non-localized U(1) Fourier analysis.
Multiresolution (multiscale) representation. Dynamics on proper
orbit/scale (inside the whole hierarchy of multiscales) in functional
space. The key ingredients are the appearance of multiscales (orbits)
and the existence of high-localized natural (eigen)modes [9]. 

\item[\bf iv)] Variational formulation (control of convergence, reductions to
algebraic systems, control of type of behaviour).
\end{itemize}

\section{SET-UP/FORMULATION}
Let us consider the following generic $\Psi DOD$
 dynamical problem 
\begin{equation}
L^j\{Op^i\}\Psi=0,
\end{equation}
described by a finite or infinite number of equations which include general classes of 
operators $Op^i$
 such as differential, integral, pseudodifferential etc 

Surely, all hierarchies and their reductions are inside this class. 

The main objects are: 

\begin{itemize}
\item[\bf i)] (Hilbert) space of states, $H=\{\Psi\}$, with a proper functional 
realization, e.g.,: $L^2$, Sobolev, Schwartz,
$C^0$, $C^k$, ... $C^\infty$, ...; 
definitely, $L^2(R^2)$, $L^2(S^2)$, 
$L^2(S^1\times S^1)$, 
$L^2(S^1\times S^1\ltimes Z_n)$
are different objects proper for different physics inside. 
E.g., two last cases describe tokamak and
stellarator, correspondingly. Of course, they are different spaces and
generate different physics. 

\item[\bf ii)] Class of smoothness. The proper choice determines natural
consideration of dynamics with/without Chaos/Fractality property.

\item[\bf iii)] Decompositions

\begin{equation}
\Psi\approx\sum_ia_ie^i
\end{equation}
via high-localized bases (wavelet families, generic wavelet packets etc), 
frames, atomic decomposition
(Fig. ~1) with the following 
main properties:
(exp) control of convergence, maximal rate of convergence  for any $\Psi$ in any $H$ [9].

\begin{figure}
\begin{center}
\begin{tabular}{c}
\includegraphics*[width=55mm]{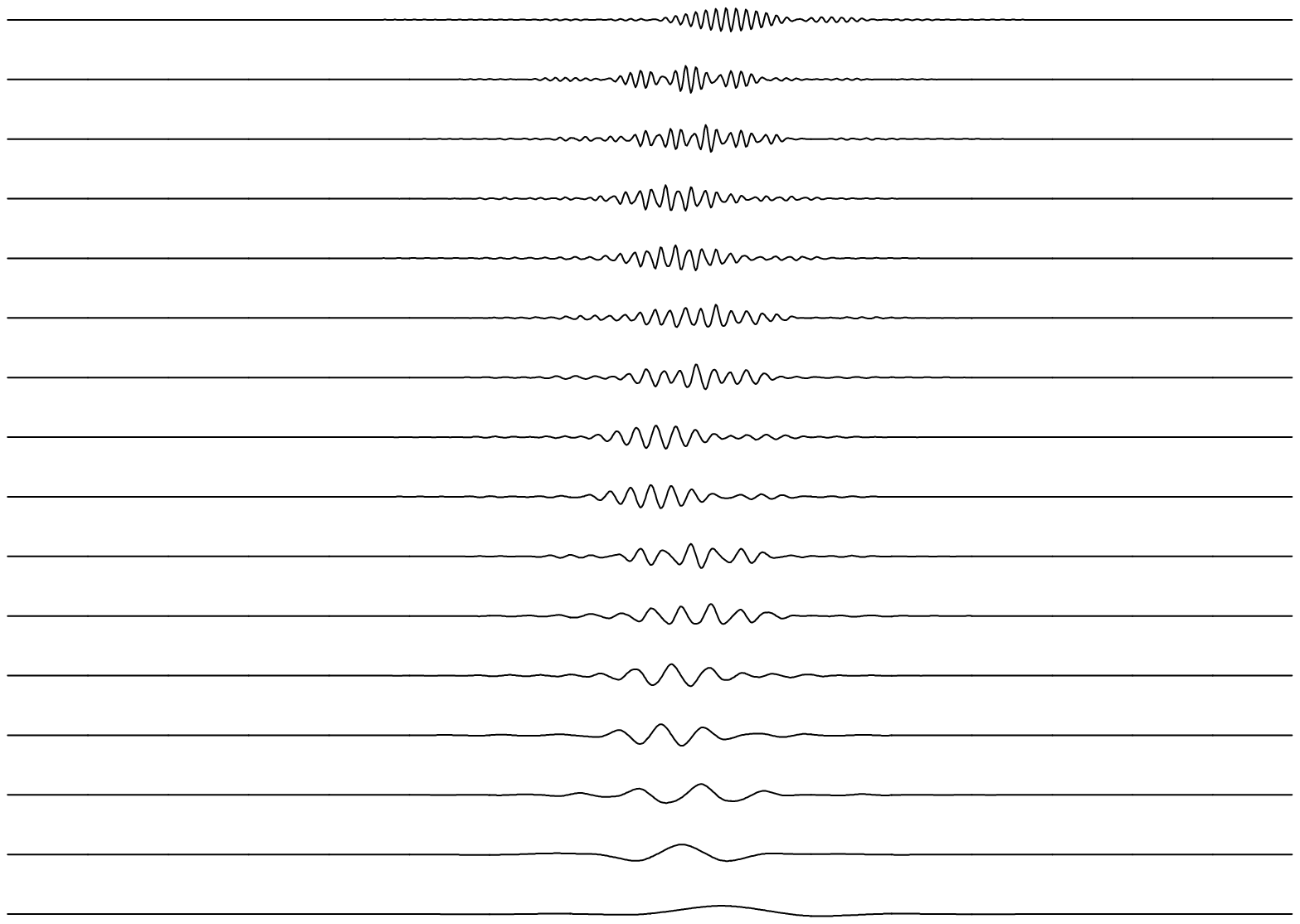}
\end{tabular}
\end{center}
\caption{Localized modes.}
\end{figure}
				
\item[\bf iv)] Observables/Operators (ODO, PDO, $\Psi DO$
, SIO,..., Microlocal analysis of Kashiwara-Shapira (with change from
functions to sheafs)) satisfy the main property - the matrix  
representation in localized bases

\begin{equation}
<\Psi|Op^i|\Psi>
\end{equation}			

is maximum sparse: 
\begin{displaymath}
\left(\begin{array}{cccc}
D_{11} & 0 &0 & \ldots\\
0    & D_{22} & 0 & \ldots\\
0    & 0    & D_{33} & \ldots\\
\vdots & \vdots & \vdots & \ddots
\end{array} \right).
\end{displaymath}

This almost diagonal structure is provided by the so-called Fast Wavelet
Transform [9].

\item[\bf v)] Measures: multifractal wavelet measures  $\{\mu_i\}$ together with the class of smoothness are very important for analysis
of complicated analytical behaviour [9]. 

\item[\bf vi)] Variational/Projection methods, from Galerkin to Rabinowitz minimax,
Floer (in symplectic case of Arnold-Weinstein curves with preservation
of Poisson or symplectic structures). Main advantages are the reduction
to algebraic systems, which provides a tool for the smart subsequent
control of behaviour and control of convergence. 

\item[\bf vii)] Multiresolution or multiscale decomposition,  $MRA$ 
 (or wavelet microscope) consists of the understanding and choosing of 
(internal) symmetry structure, e.g., affine group = \{translations, dilations\} or many 
others; construction of representation/action of this symmetry on  $H=\{\Psi\}$.

As a result of such hidden coherence together with using point vi) we'll
have: 
\begin{itemize}
\item[a).] LOCALIZED BASES  
\item[ b).] EXACT MULTISCALE DECOMPOSITION 
\end{itemize}
with the best convergence properties
and real evaluation of the rate of convergence via proper
``multi-norms''. 

Figure~ 2, 3, 5, 6 demonstrate MRA decompositions for one- and
multi-kicks while Figures 4 and 7 present the same for the case of the
generic simple fractal model, Riemann-Weierstrass function [9]. 

\item[\bf viii)] Effectiveness of proper numerics: CPU-time, HDD-space, minimal
complexity of algorithms, and (Shannon) entropy of calculations are
provided by points i)-vii) above. 

\item[\bf ix)] Quantization via $\ast$ star product or Deformation Quantization. It was considered elsewhere
[8]. 
\end{itemize}
Finally, such Variational-Multiscale approach based on points i)-ix)
provides us by the full ZOO of PATTERNS: LOCALIZED, CHAOTIC, etc. 

In next Sections we will consider details for important case of kinetic
equations. 

We present the explicit analytical construction for solutions of BBGKY
hierarchy, which is based on tensor algebra extensions of
multiresolution representation for states, observables, partitions and
variational formulation. We give explicit representation for hierarchy
of n-particle reduced distribution functions in the base of
high-localized generalized coherent (regarding underlying generic
symmetry (affine group in the simplest case)) states given by polynomial
tensor algebra of our basis functions (wavelet families, wavelet
packets), which takes into account contributions from all underlying
hidden multiscales from the coarsest scale of resolution to the finest
one to provide full information about dynamical process. In some sense,
our approach for ensembles (hierarchies) resembles Bogolyubov's one and
related approaches but we don't use any perturbation technique (like
virial expansion) or linearization procedures. Most important, that
numerical modeling in all cases shows the creation of different internal
(coherent) structures from localized modes, which are related to stable
(equilibrium) or unstable type of behaviour and corresponding pattern
(waveletons) formation.

\section{INTRODUCTION INTO PHYSICAL\\ MO\-TI\-VES}

So, we will consider the application of our numerical/analytical
technique based on local nonlinear harmonic analysis approach for the
description of complex non-equilibrium behaviour of statistical
ensembles, considered in the framework of the general BBGKY hierarchy of
kinetic equations and their different truncations/reductions. The main
points of our ideology are described below. All these facts are
well-known or described above but it is preferable to bring it together
to present our arguments in more clear form. 
\begin{itemize}

\item
Kinetic theory is an important part of general statistical physics
related to phenomena which cannot be understood on the thermodynamic or
fluid models level. 
\item
We restrict ourselves to the rational/polynomial type of nonlinearities
(with respect to the set of all dynamical variables, including
partitions) that allows to use our results, based on the so called
multiresolution framework and the variational formulation of initial
nonlinear (pseudodifferential) problems. 
\item
Our approach is based on the set of mathematical methods which give a
possibility to work with well-localized bases in functional spaces and
provide the maximum sparse forms for the general type of operators
(differential, integral, pseudodifferential) in such bases. 
\item
It provides the best possible rates of convergence and minimal
complexity of algorithms inside and, as a result, saves CPU time and HDD
space. 
\item
In all cases below by the system under consideration we mean the full
BBGKY hierarchy or some its cut-off or its various reductions. Our
scheme of cut-off for the infinite system of equations is based on some
criterion of convergence of the full solution by means of some norm
introduced in the proper functional space constructed by us. 
\item
This criterion is based on a natural norm in the proper functional
space, which takes into account (non-perturbatively) the underlying
multiscale structure of complex statistical dynamics. According to our
approach the choice of the underlying functional space is important to
understand the corresponding complex dynamics. 
\item
It is obvious that we need accurately to fix the space, where we
construct the solutions, evaluate convergence etc. and where the very
complicated infinite set of operators, appeared in the BBGKY
formulation, acts. 
\item
We underline that many concrete features of the complex dynamics are
related not only to the concrete form/class of operators/equations but
depend also on the proper choice of function spaces, where operators
act. It should be noted that the class of smoothness (related at least
to the appearance of chaotic/fractal-like type of behaviour) of the
proper functional space under consideration plays a key role in the
following. 
\item
At this stage our main goal is an attempt of classification and
construction of a possible zoo of nontrivial (meta) stable
states/patterns: high-localized (nonlinear) eigenmodes, complex
(chaotic-like or entangled) patterns, localized (stable) patterns
(waveletons). We will use it later for fusion description, modeling and
control. 
\item
Localized (meta)stable pattern (waveleton) is a good image for fusion
state in plasma (energy confinement). 
\end{itemize}

Our constructions can be applied to the following general individual
(members of ensemble under consideration) Hamiltonians:

\begin{equation}
H_N=
\sum^N_{i=1}\Big(\frac{p^2_i}{2m}+U_i(q)\Big)+
\sum_{1\leq i\leq j\leq N}U_{ij}(q_i,q_j),
\end{equation}
where the potentials
$U_i(q)=U_i(q_1,\dots,q_N)$ and $U_{ij}(q_i,q_j)$
are restricted to rational functions on the coordinates.
Let $L_s$ and $L_{ij}$
be the Liouvillean operators and 
\begin{equation}
F_N(x_1,\dots,x_N;t)
\end{equation}
be the hierarchy of  $N$-particle distribution function, satisfying the standard BBGKY 
hierarchy ($\upsilon$ is the volume): 
\begin{equation}
\frac{\partial F_s}{\partial t}+L_sF_s=
\frac{1}{\upsilon}\int d\mu_{s+1}
\sum^s_{i=1}L_{i,s+1}F_{s+1}.
\end{equation}
Our key point is the proper nonperturbative generalization of the
previous perturbative multiscale approaches. The infinite hierarchy of
distribution functions is: 
\begin{equation}
F=\{F_0,F_1(x_1;t),\dots,
F_N(x_1,\dots,x_N;t),\dots\},
\end{equation}
where 
\begin{eqnarray}
&&F_p(x_1,\dots, x_p;t)\in H^p,\\
&&H^0=R,\quad H^p=L^2(R^{6p}),\quad
F\in H^\infty=H^0\oplus H^1\oplus\dots\oplus H^p\oplus\dots\nonumber
\end{eqnarray}
with the natural Fock space like norm (guaranteeing the positivity of
the full measure): 
\begin{equation}
(F,F)=F^2_0+\sum_{i}\int F^2_i(x_1,\dots,x_i;t)\prod^i_{\ell=1}\mu_\ell.
\end{equation}

\begin{itemize}
\item
Multiresolution decomposition naturally and efficiently introduces the infinite sequence 
of the underlying hidden scales, which is a sequence of increasing closed subspaces
$V_j\in L^2(R)$:

\begin{equation}
...V_{-2}\subset V_{-1}\subset V_0\subset V_{1}\subset V_{2}\subset ...
\end{equation}

\item	
Our variational approach reduces the initial problem to the problem of
solution of functional equations at the first stage and some algebraic
problems at the second one.
As a result, the solutions of the (truncated) hierarchies have the following multiscale 
decomposition via high-localized eigenmodes ($\omega_l\sim 2^l$, $k^{s}_m\sim 2^m$)
\begin{eqnarray}
F(t,x_1,x_2,\dots)&=&
\sum_{(i,j)\in Z^2}a_{ij}U^i\otimes V^j(t,x_1,\dots),\\
V^j(t)&=&
V_N^{j,slow}(t)+\sum_{l\geq N}V^j_l(\omega_lt),\nonumber\\
U^i(x_s)
&=&U_M^{i,slow}(x_s)+\sum_{m\geq M}U^i_m(k^{s}_mx_s), \nonumber
\end{eqnarray}	
which corresponds to the full multiresolution expansion in all
underlying time/space scales. 
\item	
In this way one obtains contributions to the full solution from each
scale of resolution or each time/space scale or from each nonlinear
eigenmode. 
\item	
It should be noted that such representations give the best possible
localization properties in the corresponding (phase) space/time
coordinates. 
\item
Numerical calculations are based on compactly supported wavelets and related wavelet 
families and on evaluation of the accuracy on the level $N$
 of the corresponding cut-off of the full system w.r.t. the norm above. 
\item
Numerical modeling shows the creation of different internal structures
from localized modes, which are related to (meta)stable or unstable type
of behaviour and the corresponding patterns (waveletons) formation.
Reduced algebraic structure provides the pure algebraic control of
stability/unstability scenario. 
\item
As a final point we will consider the construction for 

CONTROLLABLE (META) STABLE WAVELETON CONFIGURATION REPRESENTING A
REASONABLE APRROXIMATION FOR THE REALIZABLE CONFINEMENT STATE.
\end{itemize}

\section{BBGKY HIERARCHY}

We start from set-up for kinetic BBGKY hierarchy and present explicit
analytical construction for solutions of hierarchy of equations, which
is based on tensor algebra extensions of multiresolution representation
and variational formulation. We give explicit representation for
hierarchy of n-particle reduced distribution functions in the base of
high-localized generalized coherent (w.r.t. underlying affine group)
states given by polynomial tensor algebra of wavelets, which takes into
account contributions from all underlying hidden multiscales from the
coarsest scale of resolution to the finest one to provide full
information about stochastic dynamical process.

Let $M$ be the phase space of ensemble of $N$ particles ($ {\rm dim}M=6N$)
 with coordinates 

\begin{eqnarray}
&&x_i=(q_i,p_i), \quad i=1,...,N,\quad
q_i=(q^1_i,q^2_i,q^3_i)\in R^3,\\
&&p_i=(p^1_i,p^2_i,p^3_i)\in R^3,
q=(q_1,\dots,q_N)\in R^{3N}.\nonumber
\end{eqnarray}

Individual and collective measures are: 

\begin{eqnarray}
\mu_i=\ud x_i=\ud q_ip_i,\quad \mu=\prod^N_{i=1}\mu_i.
\end{eqnarray}

Distribution function 
\begin{equation}
D_N(x_1,\dots,x_N;t)
\end{equation}
satisfies 
Liouville equation of motion for ensemble with Hamiltonian $H_N$ :
\begin{eqnarray}
\frac{\partial D_N}{\partial t}=\{H_N,D_N\}
\end{eqnarray}
and normalization constraint 

\begin{eqnarray}
\int D_N(x_1,\dots,x_N;t)\ud\mu=1,
\end{eqnarray}

where Poisson brackets are: 
\begin{eqnarray}
\{H_N,D_N\}=\sum^N_{i=1}\Big(\frac{\partial H_N}{\partial q_i}
\frac{\partial D_N}{\partial p_i} - \frac{\partial H_N}{\partial p_i}
\frac{\partial D_N}{\partial q_i}\Big).
\end{eqnarray}

Our constructions can be applied to the following general Hamiltonians: 
\begin{eqnarray}
H_N=\sum^N_{i=1}\Big(\frac{p^2_i}{2m}+U_i(q)\Big)+
\sum_{1\leq i\leq j\leq N}U_{ij}(q_i,q_j),  
\end{eqnarray}

where potentials 
\begin{equation}
U_i(q)=U_i(q_1,\dots,q_N)
\end{equation}
and 
\begin{equation}
U_{ij}(q_i,q_j)
\end{equation}
are not more than rational functions on coordinates.
Let $L_s$ and $L_{ij}$
 be the Liouvillean operators (vector fields) 

\begin{eqnarray}
L_s&=&\sum^s_{j=1}\Big(\frac{p_j}{m}\frac{\partial}{\partial q_j}-
\frac{\partial u_j}{\partial q}\frac{\partial}{\partial p_j}\Big)-\sum_{1\leq i\
leq j\leq s}L_{ij},\\
L_{ij}&=&\frac{\partial U_{ij}}{\partial q_i}\frac{\partial}{\partial p_i}+
\frac{\partial U_{ij}}{\partial q_j}\frac{\partial}{\partial p_j}.
\end{eqnarray}

For $s=N$ we have the following representation for Liouvillean vector field

\begin{eqnarray}
L_N=\{H_N,\cdot \}
\end{eqnarray}

and the corresponding equation of motion for ensemble:

\begin{eqnarray}
\frac{\partial D_N}{\partial t}+L_ND_N=0
\end{eqnarray}
$L_N$ is a self-adjoint operator w.r.t. standard pairing on the set of phase
space functions. Let
\begin{eqnarray}
F_N(x_1,\dots,x_N;t)=\sum_{S_N}D_N(x_1,\dots,x_N;t)
\end{eqnarray}

be the N-particle distribution function ($S_N$ is permutation group of $N$ elements). 
Then we have the hierarchy of reduced distribution functions ($V^s$
 is the corresponding normalized volume factor)
\begin{eqnarray}
F_s(x_1,\dots,x_s;t)=
V^s\int D_N(x_1,\dots,x_N;t)\prod_{s+1\leq i\leq N}\mu_i.
\end{eqnarray}

After standard manipulations we arrive to BBGKY hierarchy: 
\begin{eqnarray}
\frac{\partial F_s}{\partial t}+L_sF_s=\frac{1}{\upsilon}\int\ud\mu_{s+1}
\sum^s_{i=1}L_{i,s+1}F_{s+1}.
\end{eqnarray}

It should be noted that we may apply our approach even to more general
formulation.

For s=1,2 we have: 
\begin{eqnarray}
\frac{\partial F_1(x_1;t)}{\partial t}+\frac{p_1}{m}\frac{\partial}{\partial q_1
}
F_1(x_1;t)
=\frac{1}{\upsilon}\int\ud x_2L_{12} F_2(x_1,x_2;t)
\frac{\partial F_2(x_1,x_2;t)}{\partial t}+\\
\Big(\frac{p_1}{m}
\frac{\partial}{\partial q_1}
+\frac{p_2}{m}\frac{\partial}{\partial q_2}-L_{12}\Big)
\cdot F_2(x_1,x_2;t)
=\frac{1}{\upsilon}\int\ud x_3(L_{13}+L_{23})F_3(x_1,x_2;t)\nonumber
\end{eqnarray}
As in the general situation as in particular ones (cut-off, e.g.) we are
interested in the cases when
\begin{equation}
F_k(x_1,\dots,x_k;t)=\prod^k_{i=1}F_1(x_i;t)+G_k(x_1,\dots,x_k;t),
\end{equation}
where $G_k$ are correlators, really have additional reductions as in the simplest
case of one-particle Vlasov/Boltzmann-like systems. Then by using such
physical motivated reductions or/and during the corresponding cut-off
procedure we obtain instead of linear and pseudodifferential (in general
case) equations their finite-dimensional but nonlinear approximations
with the polynomial type of nonlinearities (more exactly,
multilinearities). Our key point in the following consideration is the
proper generalization of naive perturbative multiscale Bogolyubov's
structure.

\section{MULTISCALE ANALYSIS}

The infinite hierarchy of distribution functions satisfying BBGKY system
in the thermodynamical limit is: 
\begin{eqnarray}
F=\{F_0,F_1(x_1;t),F_2(x_1,x_2;t),\dots,\quad
F_N(x_1,\dots,x_N;t),\dots\},
\end{eqnarray}
where 

\begin{eqnarray}
F_p(x_1,\dots, x_p;t)\in H^p,\quad
H^0=R,\quad H^p=L^2(R^{6p})
\end{eqnarray}		
(or any different proper functional space), 
\begin{equation}
F\in H^\infty=H^0\oplus H^1\oplus\dots\oplus H^p\oplus\dots
\end{equation}
with the natural Fock-space like norm (of course, we keep in mind the
positivity of the full measure): 
\begin{eqnarray}
(F,F)=F^2_0+\sum_{i}\int F^2_i(x_1,\dots,x_i;t)\prod^i_{\ell=1}\mu_\ell
\end{eqnarray}
while in particular case
\begin{eqnarray}
F_k(x_1,\dots,x_k;t)=\prod^k_{i=1}F_1(x_i;t)
\end{eqnarray}
First of all we consider  $F=F(t)$ as function of the time variable only,
$F\in L^2(R)$, via multiresolution decomposition which naturally and efficiently
introduces the infinite sequence of underlying hidden scales into the
game. 

Because affine group of translations and dilations is inside the
approach, this method resembles the action of a microscope. We have
contribution to final result from each scale of resolution from the
whole infinite scale of spaces. 

Let the closed subspace $V_j (j\in {\bf Z})$  correspond to level $j$ of resolution, or to scale $j$. 
We consider a multiresolution analysis of $L^2(R)$  (of course, we may consider any different 
functional space) which is a sequence of increasing closed subspaces $V_j$ [9]:
\begin{equation}
...V_{-2}\subset V_{-1}\subset V_0\subset V_{1}\subset V_{2}\subset ...
\end{equation}
satisfying the following properties:
let $W_j$ be the orthonormal complement of $V_j$ with respect to $V_{j+1}$: 
\begin{equation}
V_{j+1}=V_j\bigoplus W_j
\end{equation}

then we have the following decomposition: 
\begin{eqnarray}
\{F(t)\}=\bigoplus_{-\infty<j<\infty} W_j
\end{eqnarray}

or  in case when $V_0$ is the coarsest scale of resolution: 

\begin{eqnarray}
\{F(t)\}=\overline{V_0\displaystyle\bigoplus^\infty_{j=0} W_j}.
\end{eqnarray}

Subgroup of translations generates basis for fixed scale number: 
\begin{equation}
{\rm span}_{k\in Z}\{2^{j/2}\Psi(2^jt-k)\}=W_j.
\end{equation}

The whole basis is generated by action of full affine group: 
\begin{eqnarray}
{\rm span}_{k\in Z, j\in Z}\{2^{j/2}\Psi(2^jt-k)\}=
{\rm span}_{k,j\in Z}\{\Psi_{j,k}\}
=\{F(t)\}.
\end{eqnarray}
Let sequence 

\begin{equation}
\{V_j^t\}, V_j^t\subset L^2(R)
\end{equation}

correspond to multiresolution analysis on time axis, 
\begin{equation}
\{V_j^{x_i}\}
\end{equation}
correspond to multiresolution analysis for coordinate $x_i$, then 

\begin{equation}
V_j^{n+1}=V^{x_1}_j\otimes\dots\otimes V^{x_n}_j\otimes  V^t_j
\end{equation}
corresponds to multiresolution 
analysis for n-particle distribution function\\ 
$F_n(x_1,\dots,x_n;t)$.
E.g., for $n=2$:

\begin{eqnarray}
V^2_0=\{f:f(x_1,x_2)=
\sum_{k_1,k_2}a_{k_1,k_2}\phi^2(x_1-k_1,x_2-k_2),\qquad
a_{k_1,k_2}\in\ell^2(Z^2)\},
\end{eqnarray}
where 

\begin{equation}
\phi^2(x_1,x_2)=\phi^1(x_1)\phi^2(x_2)=\phi^1\otimes\phi^2(x_1,x_2),
\end{equation}		
and $\phi^i(x_i)\equiv\phi(x_i)$ form a multiresolution basis corresponding to
$\{V_j^{x_i}\}$. If 
\begin{equation}
\{\phi^1(x_1-\ell)\},\ \ell\in Z
\end{equation}
form an orthonormal set, then 
\begin{equation}
\phi^2(x_1-k_1, x_2-k_2)
\end{equation}
 form an orthonormal basis for $V^2_0$.
Action of affine group provides us by multiresolution representation of
$L^2(R^2)$. After introducing detail spaces $W^2_j$, we have, e.g. 

\begin{equation}
V^2_1=V^2_0\oplus W^2_0.
\end{equation}				
Then
3-component basis for $W^2_0$ is generated by translations of three functions 
\begin{eqnarray}
&&\Psi^2_1=\phi^1(x_1)\otimes\Psi^2(x_2),\\
&&\Psi^2_2=\Psi^1(x_1)\otimes\phi^2(x_2), \nonumber\\
&&\Psi^2_3=\Psi^1(x_1)\otimes\Psi^2(x_2)\nonumber
\end{eqnarray}
Also, we may use the rectangle lattice of scales and one-dimentional
wavelet decomposition: 
\begin{equation}
f(x_1,x_2)=\sum_{i,\ell;j,k}<f,\Psi_{i,\ell}\otimes\Psi_{j,k}>
\Psi_{j,\ell}\otimes\Psi_{j,k}(x_1,x_2),
\end{equation}
where bases functions 
\begin{equation}
\Psi_{i,\ell}\otimes\Psi_{j,k}
\end{equation}
depend on
two scales $2^{-i}$ and $2^{-j}$ [9]. 

We obtain our multiscale\-/mul\-ti\-re\-so\-lu\-ti\-on 
representations below 
via the variational wavelet approach for 
the following formal representation of the BBGKY system  
(or its finite-dimensional nonlinear approximation for the 
$n$-particle distribution functions) with the corresponding obvious
constraints on the distribution functions.

\section{VARIATIONAL APPROACH}
Let $L$ be an arbitrary (non)li\-ne\-ar dif\-fe\-ren\-ti\-al\-/\-in\-teg\-ral operator 
 with matrix dimension $d$
(finite or infinite), 
which acts on some set of functions
from $L^2(\Omega^{\otimes^n})$:  
$\quad\Psi\equiv\Psi(t,x_1,x_2,\dots)=\Big(\Psi^1(t,x_1,x_2,\dots), \dots$,
$\Psi^d(t,x_1,x_2,\dots)\Big)$,
 $\quad x_i\in\Omega\subset{\bf R}^6$, $n$
 is the number of particles:  

{\setlength\arraycolsep{1pt}
\begin{eqnarray}
L\Psi&\equiv& L(Q,t,x_i)\Psi(t,x_i)=0,\\
Q&\equiv& Q_{d_0,d_1,d_2,\dots}(t,x_1,x_2,\dots,
\partial /\partial t,\partial /\partial x_1,
\partial /\partial x_2,\dots,\int \mu_k)=\nonumber\\
&=&
\sum_{i_0,i_1,i_2,\dots=1}^{d_0,d_1,d_2,\dots}
q_{i_0i_1i_2\dots}(t,x_1,x_2,\dots)
\Big(\frac{\partial}{\partial t}\Big)^{i_0}\Big(\frac{\partial}{\partial x_1}\Big)^{i_1}
\Big(\frac{\partial}{\partial x_2}\Big)^{i_2}\dots\int\mu_k.\nonumber 
\end{eqnarray}			

\noindent Let us consider now the $N$		
mode approximation for the solution as the following ansatz: 
\begin{eqnarray}
\Psi^N(t,x_1,x_2,\dots)=
\sum^N_{i_0,i_1,i_2,\dots=1}a_{i_0i_1i_2\dots}
 A_{i_0}\otimes 
B_{i_1}\otimes C_{i_2}\dots(t,x_1,x_2,\dots).
\end{eqnarray}

We shall determine the expansion coefficients from the following
conditions (different related variational approaches are considered
also): 
\begin{eqnarray}
\ell^N_{k_0,k_1,k_2,\dots}\equiv 
\int(L\Psi^N)A_{k_0}(t)B_{k_1}(x_1)C_{k_2}(x_2)\ud t\ud x_1\ud x_2\dots=0.
\end{eqnarray}
Thus, we have exactly $dN^n$ algebraical equations for  $dN^n$ unknowns 
$a_{i_0,i_1,\dots}$.
This variational ap\-proach reduces the initial problem 
to the problem of solution 
of functional equations at the first stage and 
some algebraical problems at the second.
We consider the multiresolution expansion as the second main part of our 
construction. 
The solution is parametrized by the solutions of two sets of 
reduced algebraical
problems, one is linear or nonlinear
(depending on the structure of the operator $L$)
and the rest are linear problems related to the computation of the
coefficients of the algebraic equations. These coefficients can be found
by some methods by using the compactly supported wavelet basis
functions. 

As a result the solution has the following multiscale/multiresolution
decomposition via nonlinear high-localized eigenmodes
\begin{eqnarray}
F(t,x_1,x_2,\dots)&=&\sum_{(i,j)\in Z^2}a_{ij}U^i\otimes V^j(t,x_1,x_2,\dots),\\
V^j(t)&=&V_N^{j,slow}(t)+\sum_{l\geq N}V^j_l(\omega_lt), \quad \omega_l\sim 2^l, \nonumber\\
U^i(x_s)&=&U_M^{i,slow}(x_s)+\sum_{m\geq M}U^i_m(k^{s}_mx_s), \quad k^{s}_m\sim 2^m,
 \nonumber
\end{eqnarray}

which corresponds to the full multiresolution expansion in all
underlying time/space scales. These formulae give the expansion into a
slow and fast oscillating parts. So, we may move from the coarse scales
of resolution to the finest ones for obtaining more detailed information
about the dynamical process. In this way we give contribution to our
full solution from each scale of resolution or each time/space scale or
from each nonlinear eigenmode. It should be noted that such
representations give the best possible localization properties in the
corresponding (phase)space/time coordinates. In contrast with different
approaches our formulae do not use perturbation technique or
linearization procedures. Numerical calculations are based on compactly
supported wavelets and related wavelet families and on evaluation of the
accuracy regarding norm (37):
\begin{eqnarray}
\|F^{N+1}-F^{N}\|\leq\varepsilon
\end{eqnarray}

\section{MODELING OF PATTERNS}

To summarize, the key points are: 

1. The ansatz-oriented choice of the (multidimensional) bases related to
some polynomial tensor algebra. 

2. The choice of proper variational principle. A few projection or
Galerkin-like principles for constructing (weak) solutions are
considered. The advantages of formulations related to biorthogonal
(wavelet) decomposition should be noted. 

3. The choice of bases functions in the scale spaces $W_j$ from wavelet zoo. They correspond to high-localized (nonlinear)
oscillations/excitations, nontrivial local (stable)
distributions/fluctuations, etc. Besides fast convergence properties it
should be noted minimal complexity of all underlying calculations,
especially in case of choice of wavelet packets which minimize Shannon
entropy. 

4. Operator representations providing maximum sparse representations for arbitrary 
(pseudo) differential/ integral operators: 
$\quad \ud f/\ud x$, $\ud^n f/\ud x^n$,\\ $\int T(x,y)f(y)\ud y$, etc [9]. 

5. (Multi)linearization. Besides the variation approach we can consider
also a different method to deal with (polynomial) nonlinearities:
para-products-like decompositions [9]. 

Formulae (57), (59) provide, in principle, a fast convergent
decomposition for the general solutions of the systems (31), (32), (56)
in terms of contributions from all underlying hidden internal scales. Of
course, we cannot guarantee that each concrete system (32) with fixed
coefficients will have a priori a specific type of behaviour, either
localized or chaotic. Instead, we can analyze whether such typical
structures described by qualitative definitions from Section 2.2. really
appear.

         To classify the qualitative behaviour we apply standard methods from general control 
theory or really use the control. We start from a priori unknown coefficients, the exact values 
of which will subsequently be recovered. Roughly speaking, we fix only class of nonlinearity 
(polynomial in our case) which covers a broad variety of examples of possible truncation of 
the systems. As a simple model we choose band-triangular non-sparse matrices $(a_{ij})$  in 
particular case $d=2$. These matrices provide tensor structure of bases in (extended) phase 
space and are generated by the roots of the reduced variational (Galerkin-like) systems. As a 
second step we need to restore the coefficients from these matrices by which we may classify 
the types of behaviour. We start with the two-dimensional localized base eigenmode (Fig. 9, 
d=2), which was constructed as a tensor product of the two localized one-dimensional modes 
(Fig. 8, d=1). Fig. 10, corresponding to chaotic pattern, presents the result of summation of 
series up to value of the dilation/scale parameter equal to six on the bases of symmlets with 
the corresponding matrix elements equal to one. The size of matrix is 512x512 and as a result 
we provide modeling for one-particle 
distribution function corresponding 
to standard Vlasov-like cut-off with  $F_2=F_1^2$. 
So, different possible distributions of the root values of the 
generic algebraical system (58) provide qualitatively different types of behaviour. The above 
choice provides us by a distribution with chaotic-like equidistribution. But, if we consider a 
band-like structure of matrix  $(a_{ij})$ with the band along the main diagonal with finite size 
($\ll 512$) and values, e.g. five, while the other values are equal to one, we
obtain localization in a fixed finite area of the full phase space, i.e.
almost all energy of the system is concentrated in this small volume.
This corresponds to waveleton case and is shown in Fig. 11, constructed
by means of Daubechies-based wavelet packets. Depending on the type of
solution, such localization may be present during the whole time
evolution (asymptotically-stable) or up to the needed value on time
scale, e.g., enough for plasma fusion/confinement. 

Now we discuss how to solve the inverse/synthesis problem or how to
restore the coefficients of the initial systems (31), (32). Let
\begin{equation}
L^0(Q^0)\Psi^0=0
\end{equation}

be the system (56) with the fixed coefficients $Q^0$. 
The corresponding solution $\Psi^0$
is represented by formulae (57) or (59), 
which are parametrized by roots of reduced algebraic
system (58) and constructed by some choice of the tensor product 
bases from Section 6.
The proper counterpart of the system (61) with prescribed behaviour 
$\Psi^u$, 
corresponding
to a given choice of both tensor product structure and 
coefficients $\{a_{ij}\}$ 
described above, corresponds to the class of systems like (56) but
with undetermined coefficients $Q^u$
and has the same form
\begin{equation}
L^u(Q^u)\Psi^u=0.
\end{equation}
Our goal is to restore coefficients $Q^u$ from (61), (62) and 
explicit representations for solutions 
$\Psi^0$ and $\Psi^u$. This is a standard problem in the adaptive control
theory: one adds a controlling signal
$u(x,t)$ which deforms the controlled signal $\Psi(x,t)$ from the fixed state
$\Psi^0(x,t)$ to the prescribed one $\Psi^u(x,t)$. At the same time one can 
determine the 
 parameters $Q^u$ [3]. Finally, we apply two variational constructions. The first one
gives the systems of algebraic equations for unknown coefficients,
generated by the following set of functionals

\begin{equation}
\Phi_N=\int\Big((L^0-L^u)\Psi^u_N,\Psi^0_N\Big){\rm d}\mu_N,
\end{equation}
where $N$ means the $N$-order approximation according to formulae (57). 
The unknown parameters $Q^*$ are given by $Q^*=\lim_{N\to\infty}Q^u_N$.
The second is an important additional constraint on the region 
$\mu_0$ in the phase space 
where we are interested in localization of almost all energy
$E=\int H(\Psi^u){\rm d}\mu$,
where $E$ is the proper energy Marsden-like functional [2]. 

We believe that the appearance of nontrivial localized (meta) stable
patterns observed by these methods is a general effect which present in
the full BBGKY-hierarchy, due to its complicated intrinsic multiscale
dynamics and it depends on neither the cut-off level nor the
phenomenological-like hypothesis on correlators. So, representations for
solutions like (59) and as a result the prediction of the existence of
the (asymptotically) stable localized patterns/states (waveletons) which
can realize energy confinement (fusion) states in BBGKY-like systems are
the main results of this approach. In addition, lines above in this
Section open way to solve the control problem by means of reduction from
initial (pseudodifferential) formulation to reduced set of algebraic one
(58) and as a result to create and support the needed fusion state(s).


\section{CONCLUSIONS}

Let us summarize our main results: 
\begin{itemize}
\item Physical Conjectures: 

\item{P1} 

State of fusion (confinement of energy) in plasma physics may and need
be considered from the point of view of non-equilibrium statistical
physics. According to this BBGKY framework looks naturally as first
iteration. Main dynamical variables are partitions. 

\item{P2} 

High localized nonlinear eigenmodes (Figures 1, 8, 9) are real physical
modes important for fusion modeling. Intermode multiscale interactions
create various patterns from these fundamental building blocks, and
determine the behaviour of plasma. 

High localized (meta) stable patterns (waveletons), considered as
long-living fluctuations, are proper images for plasma in fusion state
(Figures 12, 13). 

\item Mathematical framework: 

\item{M1} 
Problems under consideration, like BBGKY hierarchies (31) or their reductions (32) and 
(3)-(5), (20), (21), (26) from part II [10] are considered as  $\Psi DO$ problems in the 
framework of proper family of methods unified by effective multiresolution approach.

\item{M2} 

Formulae (59) based on generalized dispersion relations (GDR) (58)
provide exact multiscale representation for all dynamical variables
(partitions, first of all) in the basis of high-localized nonlinear
(eigen)modes. Numerical realizations in this framework are maximally
effective from the point of view of complexity of all algorithms inside.
GDR provide the way for the state control on the algbraic level. 

\item Realizability 

According to this approach, it is possible on formal level, in
principle, to control ensemble behaviour and to realize the localization
of energy (confinement state) inside the waveleton configurations
created from a few fundamentals modes only (Figures 14, 15). 

\item Open Questions 

\item{Q1} 

Definitely, all above is only very naive ensemble approach. Current
level of non-equilibrium statistical physics provides us only by BBGKY
generic framework. All related internal problems are well-known but we
still have nothing better. At the same time possible Vlasov-like
reductions or phenomenological models also look as very far from
reasonable from the point of view of the fusion problem set-up. 

\item{Q2} 

Considering for allusion successful areas of physics like
superconductivity, for example, we may conclude that only microscopic
BCS formulation provides the full explanation although Ginzburg-Landau
(GL) phenomenological approach and even Froelich's and London's ones
contributed to the general picture. Whether Vlasov equations are the
analogue of GL ones and whether it is possible to construct microscopic
model for plasma, these two important questions remain unanswered at
present time. 

\item{Q3} 

It may be natural also that approaches proposed in this paper and
related ones are wrong because the proper and adequate framework for
solution of fusion problem is related to confinement of magnetic lines
or loops (new physical dynamical variables instead of partitions) or
fluxes instead of confinement of localized point modes (attribute of any
local field theory) considered as new and really proper physical
variables. Such approach demands the topological background related to
proper mathematical constructions. 
As allusion it is possible to consider the description of (fractional) quantum Hall effect by 
means of Chern-Simons/anyon models which allow to describe the dynamics on (of) knots and 
braids analytically. Anyway, it is still possible to apply successfull methods from (M1) 
and (M2) here too. Remarks in Section 2.2. demonstrate interesting relations. Other open 
possibility is related to taking into account internal quantum properties. From this point of 
view our approach is very useful because we unify quantum description and its classical 
counterpart in the general $\Psi DO$  framework.
\end{itemize}

\section{ACKNOWLEDGEMENTS}

We are very grateful to Professors E. Panarella (Chairman of the
Steering Committee), R. Kirkpatrick (LANL) and R.E.H. Clark, G. Mank and
his Colleagues from IAEA (Vienna) for their help, support and kind
attention before, during and after 6th Symposium on Fusion Research
(March 2005, Washington, D.C.). We are grateful to Dr. A.G. Sergeyev for
his permanent support in problems related to hard- and software.

\begin{twocolumn}

\begin{figure}
\begin{center}
\begin{tabular}{c}
\includegraphics[width=55mm]{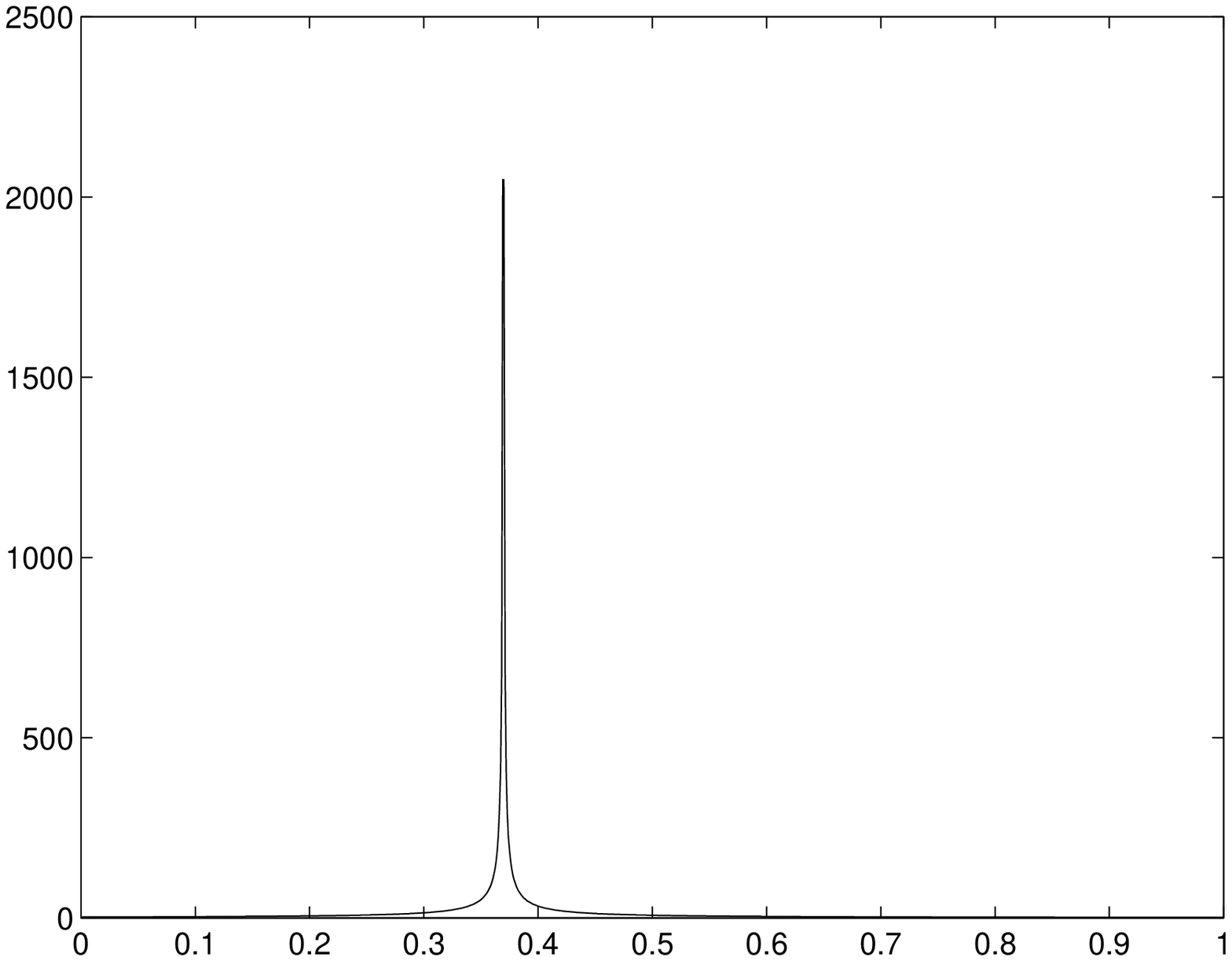}
\end{tabular}
\end{center}
\caption{Kick.}
\end{figure}

\begin{figure}
\begin{center}
\begin{tabular}{c}
\includegraphics[width=55mm]{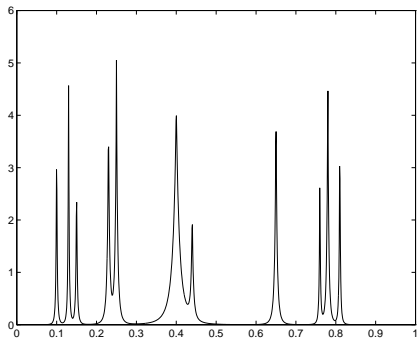}
\end{tabular}
\end{center}
\caption{Multi-Kicks.}
\end{figure}

\begin{figure}
\begin{center}
\begin{tabular}{c}
\includegraphics[width=55mm]{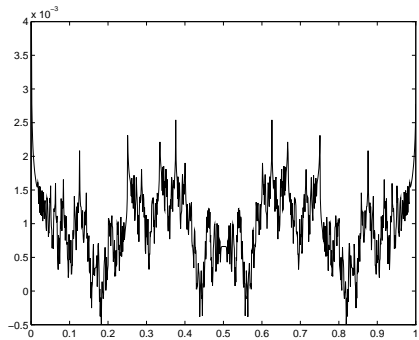}
\end{tabular}
\end{center}
\caption{RW-fractal.}
\end{figure}

\begin{figure}
\begin{center}
\begin{tabular}{c}
\includegraphics[width=55mm]{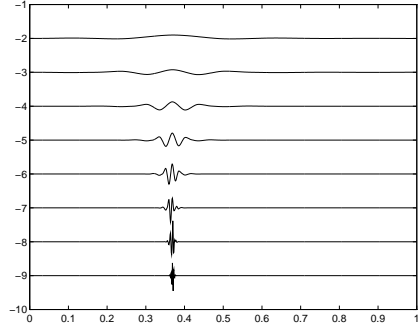}
\end{tabular}
\end{center}
\caption{MRA for Kick.}
\end{figure}

\begin{figure}
\begin{center}
\begin{tabular}{c}
\includegraphics[width=55mm]{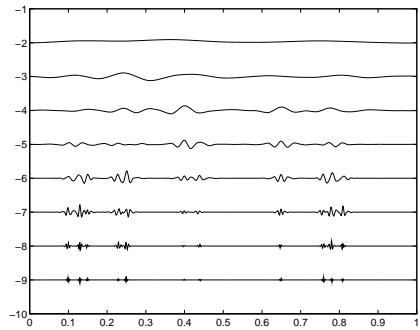}
\end{tabular}
\end{center}
\caption{MRA for Multi-Kicks.}
\end{figure}

\begin{figure}
\begin{center}
\begin{tabular}{c}
\includegraphics[width=55mm]{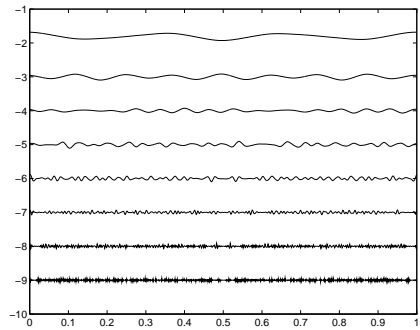}
\end{tabular}
\end{center}
\caption{MRA for RW-fractal.}
\end{figure}

\end{twocolumn}

\newpage

\onecolumn

\begin{figure}
\centering
\includegraphics*[width=70mm]{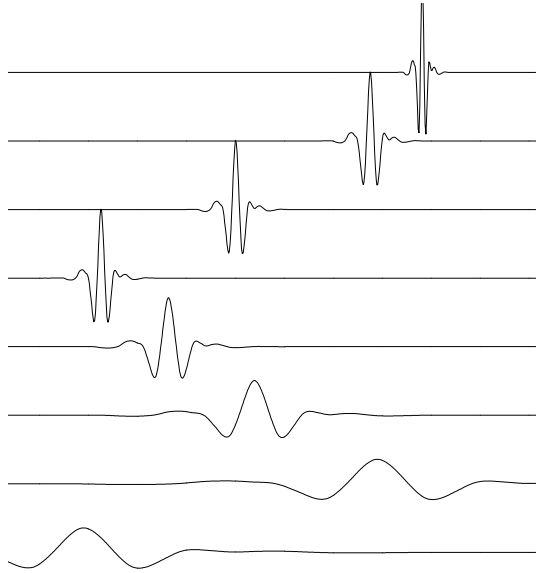} 
\caption{Localized one-dimensional modes.}                                                                       
\end{figure}                                                                                     

\begin{figure}[htb]                                                                    
\centering                                                                             
\includegraphics[width=100mm]{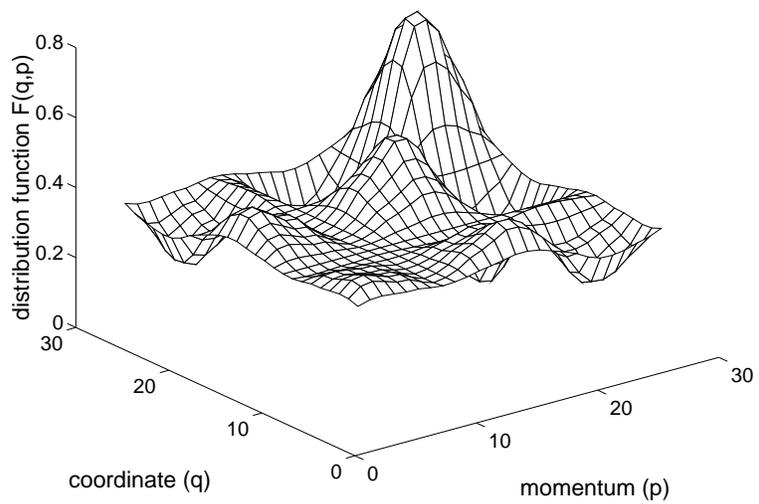}                                         
\caption{Localized mode contribution to distribution 
function.}                                                        
\end{figure}

\newpage

\begin{figure}[htb]
\centering
\includegraphics[width=100mm]{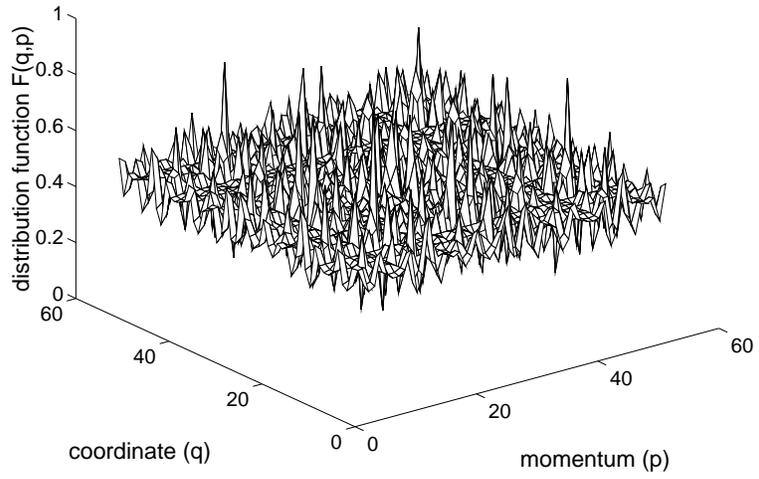}
\caption{Chaotic partition.}
\end{figure}

\begin{figure}
\begin{center}
\begin{tabular}{c}                                                                                    
\includegraphics[width=100mm]{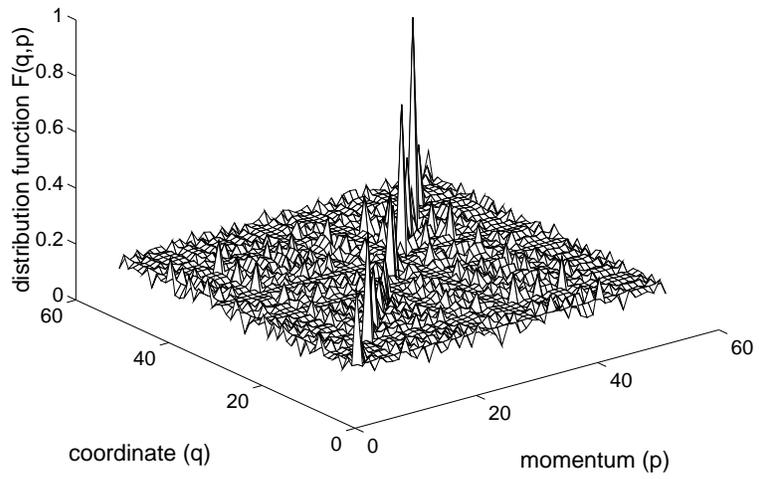} 
\end{tabular}
\end{center}                                                         
\caption{Localized partition.}                                                          
\end{figure}


\newpage

\begin{figure}[htb]
\centering 
\includegraphics[width=100mm]{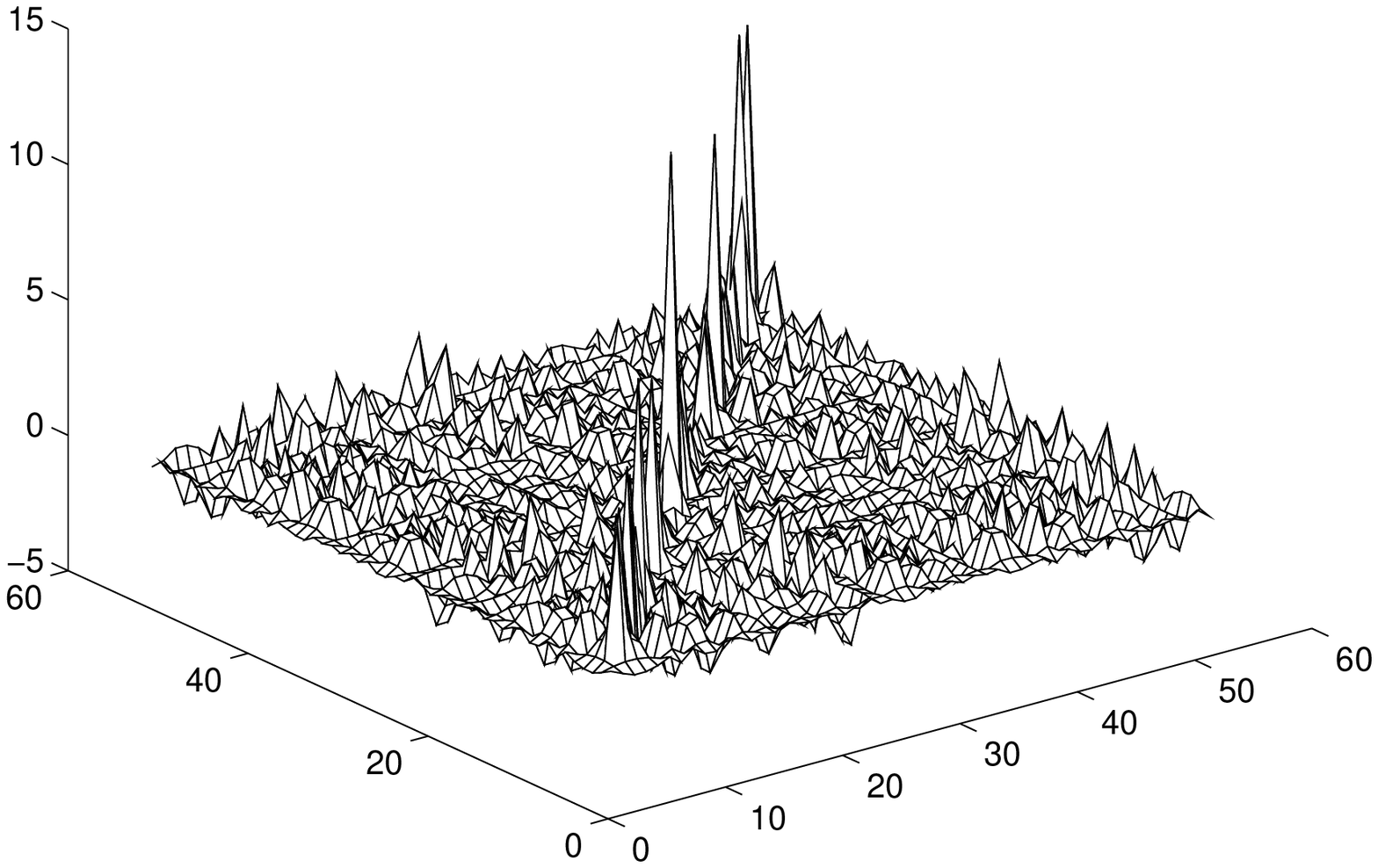} 
\caption{Fusion-like state.}
\end{figure}

\begin{figure}[htb]
\centering 
\includegraphics[width=100mm]{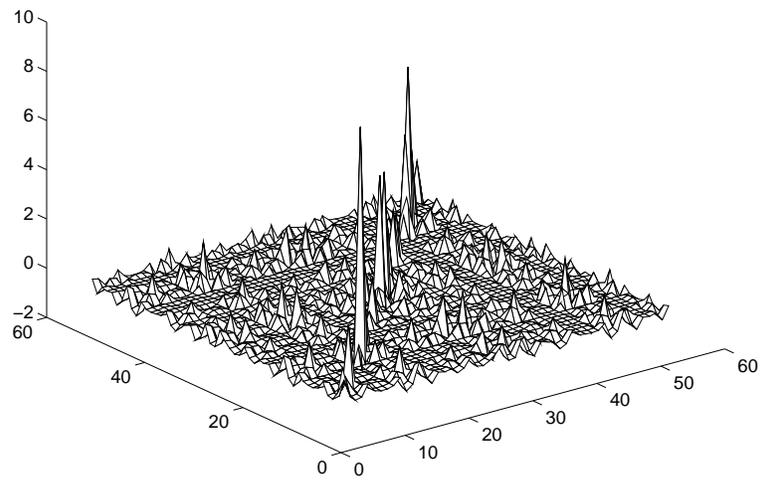} 
\caption{Fusion-like state: waveleton mode.}
\end{figure}


\begin{figure}[htb]
\centering 
\includegraphics[width=100mm]{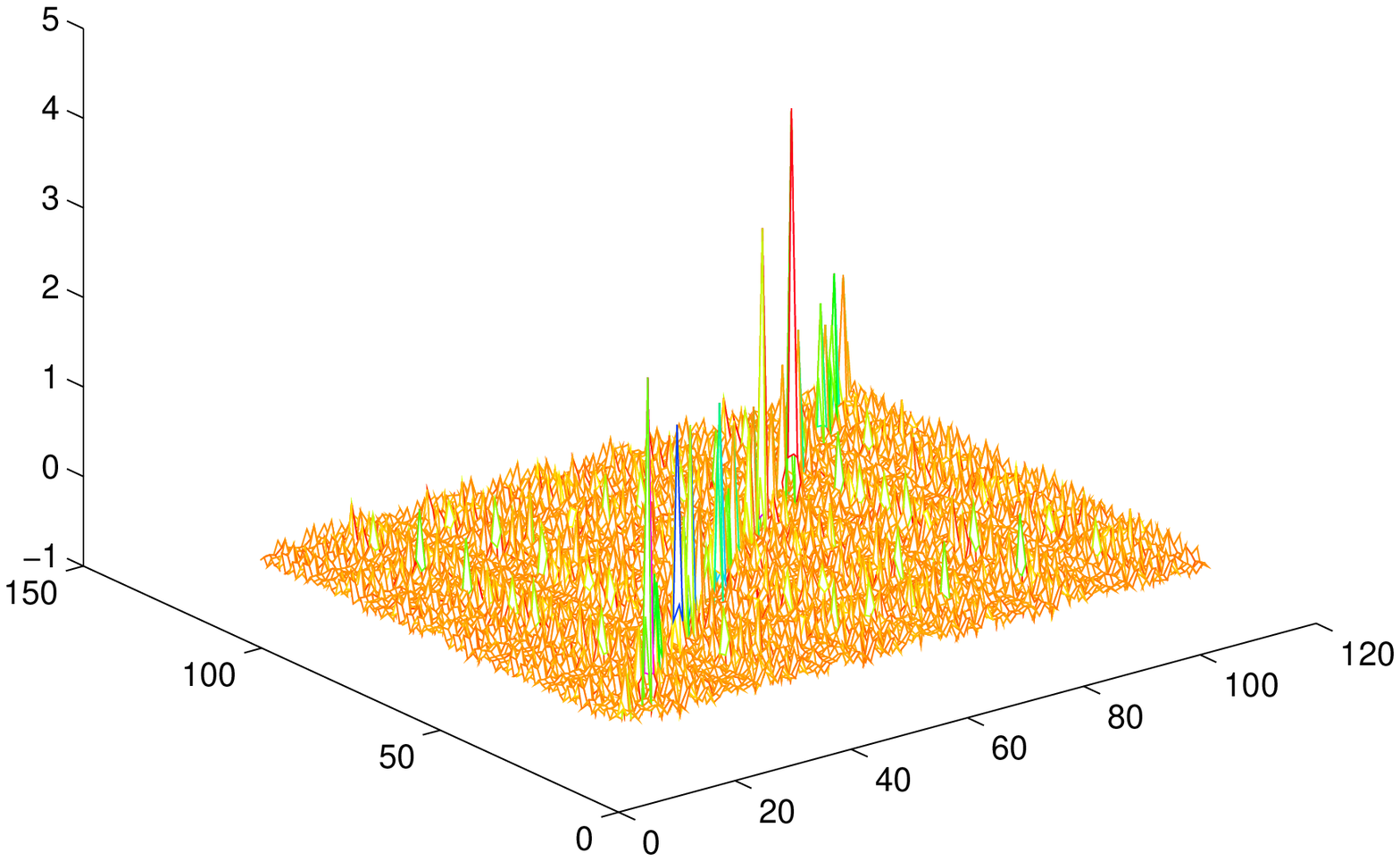} 
\caption{Fusion-like state.}
\end{figure}

\begin{figure}[htb]
\centering 
\includegraphics[width=100mm]{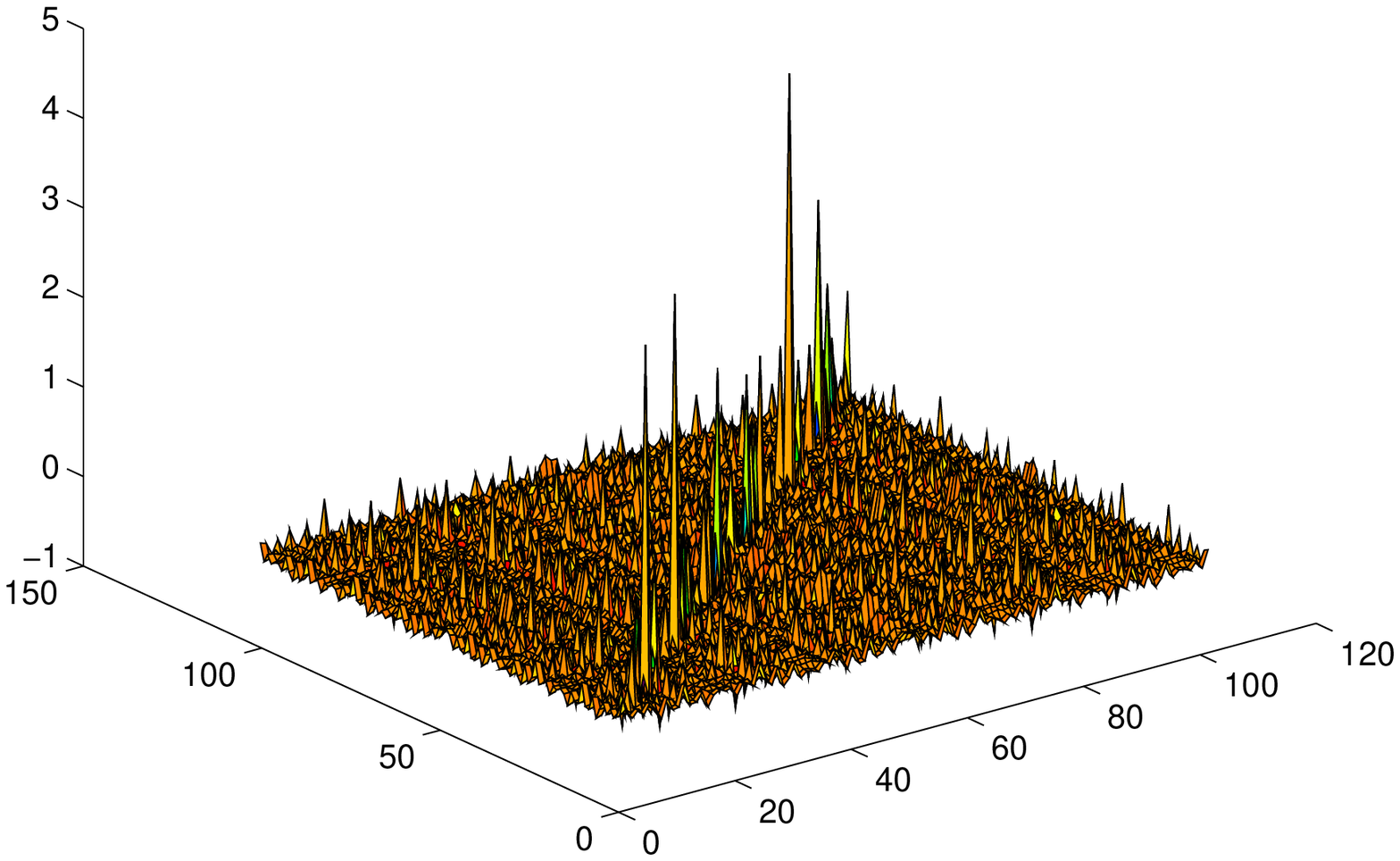} 
\caption{Fusion-like state.}
\end{figure}


\end{document}